\documentclass[letter]{aa}  

\usepackage{graphicx}
\usepackage{lscape}
\usepackage{txfonts}
\begin{document}

   \title{V363\,Cas: a new lithium rich Galactic Cepheid.\thanks{Based on observations made with the Italian Telescopio Nazionale Galileo (TNG) operated by the Fundación Galileo Galilei (FGG) of the Istituto Nazionale di Astrofisica (INAF) at the Observatorio del Roque de los Muchachos (La Palma, Canary Islands, Spain).}}

%   \subtitle{I. Overviewing the $\kappa$-mechanism}

\author{G. Catanzaro \inst{1}
          \and
          V. Ripepi \inst{2} 
          \and 
          G. Clementini \inst{3}
          \and
          F. Cusano \inst{2}
          \and
          G. De Somma \inst{2}
          \and
          S. Leccia \inst{2}
          \and
          M. Marconi \inst{2}
          \and \\
          R. Molinaro \inst{2}
          \and
          M. I. Moretti \inst{2}
          \and
          I. Musella \inst{2}
          \and
          V. Testa \inst{4}
}

\institute{ INAF-Osservatorio Astrofisico di Catania, Via S.Sofia 78, 95123, Catania, Italy \\
             \email{giovanni.catanzaro@inaf.it}
             \and
INAF-Osservatorio Astronomico di Capodimonte, Salita Moiariello 16, 80131, Naples, Italy\\  \email{vincenzo.ripepi@inaf.it}
             \and
INAF-Osservatorio di Astrofisica e Scienza dello Spazio, Via Gobetti 93/3, I-40129 Bologna, Italy\\ 
             \and
             INAF – Osservatorio Astronomico di Roma, via Frascati 33, I-00078 Monte Porzio Catone, Italy
             }

   \date{}

% \abstract{}{}{}{}{} 
% 5 {} token are mandatory
 
  \abstract
   {Classical Cepheids (DCEPs) are important astrophysical objects not only as standard candles in the determination of the cosmic distance ladder, but also as a testbed for the stellar evolution theory, thanks to the strict connection between their pulsation [period(s), amplitudes]  and stellar (luminosity, mass, effective temperature, metallicity) parameters.}   
   {We aim at unveiling the nature of the Galactic DCEP V363\,Cas and other DCEPs showing cosmic abundances of lithium in their atmospheres.}
   {We have collected three epochs high-resolution spectroscopy for  V363\,Cas with HARPS-N@TNG. Accurate stellar parameters: effective temperatures, gravities, microturbulences, radial velocities, and metal abundances were measured for this star.}
   {We detected a lithium abundance of  A(Li)\,=\,2.86\,$\pm$\,0.10~dex, along with  iron, carbon and oxygen abundances of  [Fe/H]\,=\,$-$0.30\,$\pm$\,0.12~dex, [C/H]\,=\,$-$0.06\,$\pm$\,0.15~dex and [O/H]\,=\,0.00\,$\pm$\,0.12~dex. V363 Cas is the fifth among the Milky Way DCEPs to exhibit a Li-rich feature. An analysis of historical time-series spanning a hundred  year interval shows  that the period of V363\,Cas is increasing, with a sharp acceleration after HJD= 2453000. This is a clear hint of first crossing of the instability strip.}
   {Our results favour the scenario in which the five Galactic Li-rich DCEPs are first-crossing the instability strip having had slowly-rotating progenitors during their main sequence phase.}


   \maketitle
%
%-------------------------------------------------------------------
\section{Introduction}

%Classical Cepheids (DCEPs) are the most important primary distance indicators for the extragalactic distance scale thanks to their high instrinsic luminosity ($M_V \sim -2-9)$),  large pulsation amplitude (0.1-1 mag) and above all to their characteristic period-luminosity relation \citep{Leavitt1912} that allow to estimate their distance on the basis of the measurement of the period only. They are also fundamental tracers of the young stellar population (ages 50-500 Myrs) and are usually used to study the metallicity gradient of the Milky Way (MW) disk \citep[see e.g.][and references therein]{Genovali2015} or trace the tri-dimensional structure of the host system \citep[e.g.][and references therein]{Ripepi2017,Skowron2019}. But DCEPs are also important for the stellar evolutionary theory, thanks to period-density relation that allows to link the pulsational properties of these stars to their intrinsic 

Among Classical Cepheids (DCEPs), the primary Population~I standard candle within the Local Group, the shortest period DCEPs are likely associated with the first crossing of the instability strip (IS) and are expected to show peculiar chemical features, if  compared with canonical, more evolved blue loop pulsators.

In this letter we report the discovery of a new rare lithium-rich DCEP. Only four DCEPs showing enhanced lithium abundance (via detection of the \ion{Li}{I} 6707.766 \AA) have been discovered in the Galaxy so far \citep{luck11,Kovtyukh2016,Kovtyukh2019} and an additional one was detected in the Large Magellanic Cloud \citep[LMC,][]{luck92}. All these objects show a lithium abundance  A(Li)$\sim$3.0 dex, in contrast to the majority of the Galactic DCEPs,  which show A(Li)$<$1.2 dex \citep{luck11}. This discovery was surprising as Li is expected to be depleted by proton-capture after the first dredge-up (1DU) occurring at the beginning of the Red Giant Branch (RGB) phase \citep{Iben1967}. A natural explanation %of this phenomenon 
is that these  DCEPs are at their first crossing of the IS and their envelopes do not show the signature of nuclear processes occurred during the Main Sequence (MS) phase. Indeed, according to \citet{Kovtyukh2019}, at least three of out of the four Milky Way (MW) Li-rich DCEPs also show abundances of the CNO species which are consistent with the solar values, i.e. not processed by the CN-cycle. However,  the 1DU is not the only phenomenon capable of depleting lithium. Rotational mixing  can in fact reduce the lithium abundance by a factor of a hundred in a fraction of the MS lifetime, for sufficiently fast rotating MS stars \citep[e.g.][]{Brott2011}. This would then explain the scant number of Li-rich DCEPs. Indeed, as noted by \citet{Kovtyukh2019}, about 80\% of the DCEPs that are expected to be at their first crossing (about 5\% of the total) are Li-depleted. Therefore it can be hypothesized that the progenitors of the Li-rich and Li-depleted DCEPs (B stars) were, respectively, slow and fast rotating stars, when on the MS. It is known that a fraction ($\sim$15\%) of the B-stars show $v sin i < 20$ km/s \citep{Huang2010}, while  their large majority rotates much faster. It is thought that the slow rotators loose most of their angular momentum on the MS due to stellar winds enhanced by the rotation itself \citep{Maeder2000}, hence when they become DCEPs they show the moderate rotational velocities typical of these stars. \par
An additional feature of the Li-rich DCEPs is that they most frequently are  multi-mode pulsators. Among the four MW Li-rich DCEPs, ASAS J075842-2536.1 and ASAS J131714-6605.0 both pulsate in the first and second overtone (DCEP\_1O2O), V371 Per pulsates in the fundamental and first overtone (DCEP\_F1O), whereas V1033\,Cyg is only a fundamental mode (DCEP\_F) pulsator. 
According to  \citet{Kovtyukh2019}, multi-mode DCEPs have a less efficient mixing in their envelope than DCEP\_F, hence would preferentially tend to retain their Li. 

Even if other more complex processes can address the presence of lithium in DCEPs \citep[see][for a detailed discussion]{Kovtyukh2019}, the basic mechanism to explain Li-rich DCEPs is their passage through the IS at the first crossing. This occurrence can be verified  by measuring the rate of period change due to  evolution along the Hertzsprung-Russell Diagram (HRD), as the period is expected to increase at the first and third crossing while decreasing at the second one \citep[see e.g.][]{Turner2006}. The data available to date allowed \cite{Kovtyukh2019} to detect a quick period change in  V1033\,Cyg whereas they were  insufficient to detect period changes in the other three MW DCEPs.  

In the course of a large project devoted to  measure the chemical abundance of a hundred un-characterized or newly discovered Galactic DCEPs, we obtained high-resolution spectroscopy for V363\,Cas and discovered the presence of a deep \ion{Li}{I} 6707.766 {\AA} line in the spectra of this MW DCEP. 
V363\,Cas has long been considered a fundamental mode (ab-type) RR Lyrae variable with period $P\sim0.546$ days \citep[e.g.][]{Nowakowski1988}. However,  \citet{Hajdu2009} showed  that the star is in fact a multi-mode DCEP pulsating in the first (1O) and second (2O) overtone modes, with a period ratio of P2/P1$\sim$0.801, which is typical of such multimode DCEPs \citep[see, e.g. Fig. 3 in][]{Udalski2018}. Nevertheless, a number of  recent papers still erroneously considered    V363\,Cas as an RR Lyrae star \citep[e.g.][]{Dambis2013,Prudil2020}. Furthermore, \citet{Kervella2019a,Kervella2019b} claimed that V363\,Cas is a binary RR Lyrae with an upper limit 0.2\,M$_\odot$ low mass companion.

In the following we confirm \citet{Hajdu2009}  classification of V363\,Cas as a multi-mode DCEP, based on both the star position on the HRD and an analysis of the photometric time-series data available for the star.

%The paper is organized as follows: in Sect~\ref{sect:spect} we discuss the spectroscopic observations and the data analysis performed to carry out fundamental astrophysical quantities such as: T$_{\rm eff}$, $\log g$, microturbulent, rotational and radial velocities, as well as lithium and metal abundances, in Sect.~\ref{sect:phot} we report on the photometric analysis, and finally in Sect.~\ref{sect:discus} we discuss our results in view of the position of our target in the HRD.

\section{Spectroscopic observations and data analysis}
\label{sect:spect}
Multiphase spectroscopic observations of %the Galactic Classical DCEP
V363\,Cas were obtained at the 3.5m Telescopio Nazionale Galileo (TNG) equipped with the HARPS-N 
instrument, in three nights, November 21, 26 and December 19, 2019. HARPS-N features an echelle spectrograph covering the wavelength range between 3830 to 6930 {\AA}, with a spectral resolution R=115,000. The signal-to-noise ratio (SNR) varies from 50 to 100 at $\lambda$ = 5000 {\AA}. Main characteristics of V363 Cas are summarised in Table~\ref{tab:star}.

Reduction of all the spectra, which included  bias subtraction, spectrum extraction, flat fielding and wavelength calibration,  was performed using the HARPS reduction pipeline. 
Radial velocities were measured by cross-correlating each spectrum with a synthetic template, using the IRAF task {\it FXCOR} and excluding 
Balmer lines as well as wavelength ranges  containing telluric lines. The IRAF package RVCORRECT was adopted to determine the heliocentric velocity, by  correcting the spectra for the Earth’s motion.

\begin{table}
\centering
\caption{Main characteristics of V363\,Cas. $<V>$ is the intensity--averaged magnitude of the star.}
\label{tab:star}
\setlength{\tabcolsep}{3.5pt}
\begin{tabular}{ccccc}
\hline
\hline            
\noalign{\smallskip}
{\it Gaia} ID & l & b  & $<V>$ & $E(B-V)$ \\
\noalign{\smallskip}
       & \degr & \degr & mag & mag \\
\noalign{\smallskip}
\hline            
\noalign{\smallskip}       
429162271910068352 & 118.46 & $-$02.217 & 10.550 & 0.437$^{a}$ \\

\hline                                      \end{tabular}
\tablefoot{~a =  \citet{Kervella2019b}}
\end{table}

To measure the elemental abundances we first need to estimate main stellar atmospheric parameters such as the effective temperature (T$_{\rm eff}$), surface gravity ($\log g$), microturbulent ($\xi$) and the total lines broadening, measured in our spectra equal to 15\,$\pm$\,1~km s$^{-1}$. 
\begin{figure}
\centering
\includegraphics[width=8.5cm]{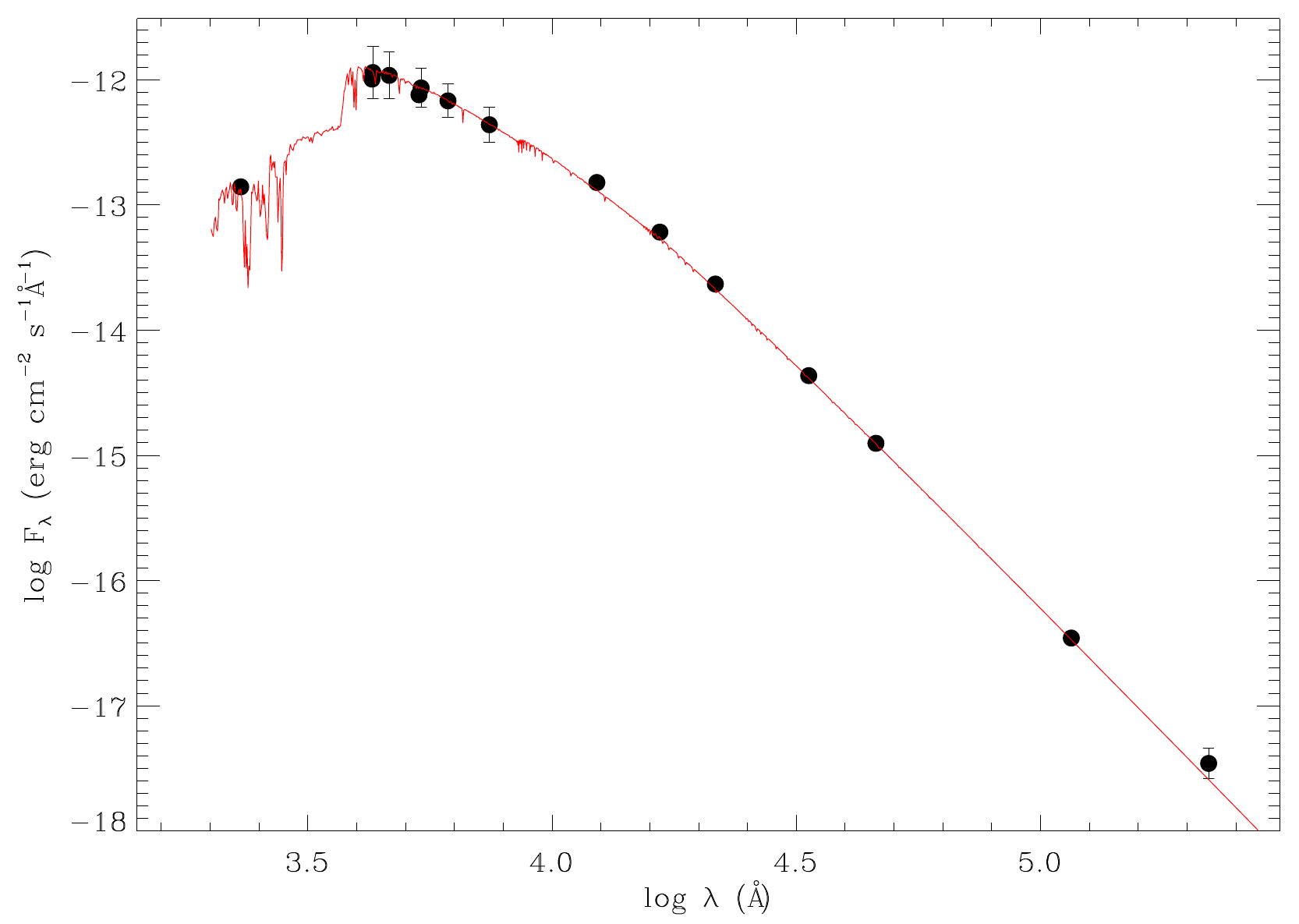}
\caption{Spectral energy distribution of V363\,Cas. Filled dots represent the observed fluxes as retrieved from the VOSA tool. A red line shows  the theoretical flux computed using the ATLAS9 model for T$_{\rm eff}$\,=\,6660~K and $\log g$\,=\,2.0.}
\label{sed}
\end{figure}
\begin{table*}
\centering
\caption{Atmospheric parameters. For each spectrum of V363\,Cas we list: heliocentric julian date (HJD) at mid exposure (column 1), pulsation phase (column 2), effective temperature (column 3), gravity (column 4), microturbulent and radial velocities (columns 5 and 6), iron and lithium abundances (column 7 and 8) expressed in a logarithmic scale relative to hydrogen.}
\label{table_sum_spectro}
\begin{tabular}{ccccccccc}
\hline \hline
      HJD  & Phase & T$_{\rm eff}$ & $\log g$ &    $\xi$      & v$_{\rm rad}$     & A(Fe) & A(Li) \\
 2400000+  &       & (K)           &          & (km s$^{-1}$) & (km s$^{-1}$)     &       &  \\
 \hline                                                                                                                
 58809.5196 & 0.996 & 6650 $\pm$ 170 & 1.9 $\pm$ 0.2 & 2.1 $\pm$ 0.6 & $-$55.2 $\pm$ 0.1 & 7.24 $\pm$ 0.12 & 2.86 $\pm$ 0.10 \\
 58814.4665 & 0.046 & 6620 $\pm$ 180 & 2.0 $\pm$ 0.2 & 2.4 $\pm$ 0.6 & $-$55.9 $\pm$ 0.1 & 7.19 $\pm$ 0.12 & 2.86 $\pm$ 0.10 \\
 58837.3996 & 0.002 & 6710 $\pm$ 160 & 2.2 $\pm$ 0.3 & 2.1 $\pm$ 0.5 & $-$50.2 $\pm$ 0.1 & 7.28 $\pm$ 0.12 & 2.86 $\pm$ 0.10 \\
\hline                    
\end{tabular}
\end{table*}
A T$_{\rm eff}$ value for each spectrum of V363\,Cas  was estimated using the line depth ratios (LDRs) method \citep{Kovtyukh2000}, which is commonly used in the literature for DCEPs. We measured about 32 LDRs in each spectrum. An iterative procedure was then applied to determine the microturbulent velocity $\xi$, iron  abundance [A(Fe)] and surface gravity ($\log g$). $\xi$ values were estimated by demanding  the slope of the iron abundance as a function of the equivalent width (EW) to be null, that is, the iron abundance to not depend on EWs. To this purpose we used a sample of 145 \ion{Fe}{I} spectral lines, extracted from the line list of  %published by
\citet{Romaniello2008}. The iron content was  estimated by converting the measured EWs into abundances through the WIDTH9 code  \citep{kur81}, after generating an appropriate model atmosphere with the ATLAS9 LTE code \citep{kur93,kur93b}. At this stage, we neglected the $\log g$, as the \ion{Fe}{I} lines are insensitive to this parameter. EWs were measured using an {\it IDL}\footnote{IDL (Interactive Data Language) is a registered trademark of Harris Geospatial Solutions} semi-automatic custom routine which allowed us to minimize errors in the continuum evaluation on the wings of the spectral lines.
Then, we estimated the surface gravity by imposing the ionization balance between \ion{Fe}{I} and \ion{Fe}{II} lines. For the \ion{Fe}{II}, we used a list of 24 lines extracted from the compilation by \citet{Romaniello2008}. The atmospheric parameters derived for each spectrum  of V363\,Cas  are summarized in Table~\ref{table_sum_spectro}. Since we obtained consistent temperatures and gravities for each night, we adopted the weighted average values, i.e. T$_{\rm eff}$\,=\,6660\,$\pm$100~K and $\log g$\,=\,2.0\,$\pm$\,0.1, to reproduce the observed spectral energy distribution (SED) with the synthetic flux computed using the ATLAS9 code. The observed flux was retrived from the VOSA tool \citep{bayo08} and corrected for reddening adopting $E(B-V)$\,=\,0.437 mag  \citep{Kervella2019b} and the \citet{fitz1999} extinction law. 
The comparison between observed and the theoretical SEDs is shown in Fig.~\ref{sed}. Furthermore, using the distance inferred from the {\it Gaia} DR2 parallax \citep[$\pi$\,=\,0.7669\,$\pm$\,0.0278 mas, to which we applied a zero point correction of 0.049 mas][]{Groenewegen2018}, we derived a bolometric luminosity of  $L/L_\odot$\,=\,260\,$\pm$\,49 for V363\,Cas.

\begin{figure*}
\centering
\includegraphics[width=13cm]{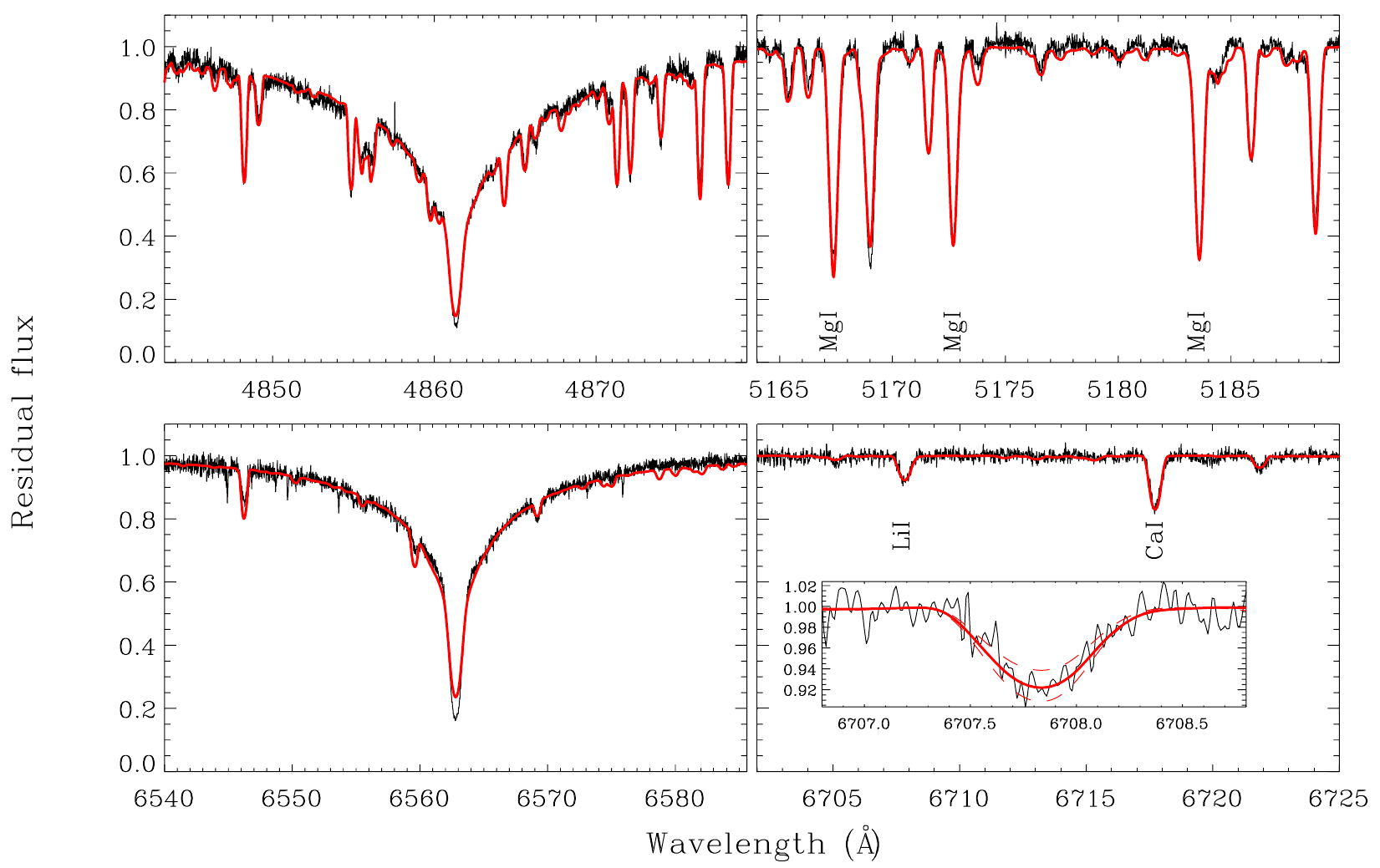}
\caption{Comparison between the spectrum of  V363\,Cas obtained on November 26, 2019 (black line) and the synthetic spectrum (red line) in four main spectral intervals centered, respectively, on H$\beta$ (top-left panel), the  \ion{Mg}{I} triplet at $\lambda \lambda$5167.3216, 5172.6843, and  5183.6042 {\AA} (top-right panel), H$\alpha$ (bottom-left panel) and the \ion{Li}{I} line at 6707.766 {\AA} (bottom-right panel). In the insert, a solid red line shows the fit of the lithium line, while dashed lines represent limits for the experimental error ($\delta$\,=\,$\pm$\,0.1~dex).}
\label{spectra}
\end{figure*}

The atmospheric parameters listed in Table~\ref{table_sum_spectro} were used as inputs for the abundance analysis which was performed following the procedures in \citet{catanzaro19}. For the analysis we used the spectrum acquired on November, 26 2019 which is the one with the highest SNR (up to 100). The abundances of the 26 species we detected in this spectrum are provided  in Table~\ref{table_abund}. In Fig.~\ref{spectra} we show the comparison between observed and synthetic spectra in four main spectral regions, namely, H$\beta$ and H$\alpha$ (as a check for the effective temperature), the \ion{Mg}{I} triplet (as a check for $\log g$), and the  \ion{Li}{I} 6708.766 {\AA} line. For the synthesis of the \ion{Li}{I} line, we took into account the hyperfine structure and the close \ion{Fe}{I} 6708.282 {\AA} line.

As a general trend, V363\,Cas appears to be slightly metal poor, since the iron peak elements show underabundances with respect to the Sun composition \citep{grevesse10}. The  low iron content, [Fe/H]\,$\approx$\,$-$0.30, is consistent with a previous estimate by \citet{fernley97} ($-$0.38). We paid  particular attention to light elements involved in the first dredge up mixing, such as carbon, oxygen, and sodium\footnote{Unfortunately, we did not observe any N spectral lines in our spectral range.}. Carbon and oxygen abundances are in agreement with the solar values, while sodium is slightly under-abundant. The lithium line was  reproduced with an abundance of A(Li)\,=\,2.86\,$\pm$\,0.10. This value is in good agreement with results from the standard big bang nucleosynthesis theory, which predicts a lithium  abundance of A(Li)\,=\,2.72\,$\pm$\,0.06 dex \citep{Cyburt2008}.

\begin{table}
\centering
\caption{Elemental abundances of V363\,Cas,  expressed in terms of the solar  abundances  \citep{grevesse10},  for  26 chemical species we  measured in our target. Columns labelled with N represents number of lines used in the analysis. Results were obtained from  the spectrum acquired on November 26, 2019.}
\label{table_abund}
\begin{tabular}{lrrlrr}
\hline
\hline            \noalign{\smallskip}
            
El  &  [El/H]~~~~~& N & El & [El/H]~~~~~ & N \\

            \noalign{\smallskip}
\hline
            \noalign{\smallskip}

Li  &     1.76 $\pm$ 0.14 &  1 & Mn  & $-$0.55 $\pm$ 0.12  & 8 \\
C   &  $-$0.06 $\pm$ 0.15 &  4 & Fe  & $-$0.30 $\pm$ 0.12  &169 \\
O   &     0.00 $\pm$ 0.12 &  2 & Ni  & $-$0.08 $\pm$ 0.11  & 14 \\
Na  &  $-$0.11 $\pm$ 0.12 &  4 & Cu  & $-$0.40 $\pm$ 0.10  & 2 \\
Mg  &  $-$0.06 $\pm$ 0.12 &  5 & Zn  & $-$0.22 $\pm$ 0.15  & 3 \\
Al  &  $-$0.38 $\pm$ 0.15 &  2 & Sr  &    0.16 $\pm$ 0.10  & 2 \\
Si  &  $-$0.08 $\pm$ 0.14 & 10 & Y   &    0.15 $\pm$ 0.10  & 3 \\
S   &     0.06 $\pm$ 0.16 &  3 & Zr  &    0.08 $\pm$ 0.15  & 3 \\
Ca  &     0.00 $\pm$ 0.15 &  4 & Ba  &    0.86 $\pm$ 0.16  & 5 \\
Sc  &  $-$0.08 $\pm$ 0.10 &  3 & La  &    0.33 $\pm$ 0.15  & 2 \\
Ti  &  $-$0.06 $\pm$ 0.15 & 20 & Ce  &    0.05 $\pm$ 0.15  & 1 \\
V   &  $-$0.49 $\pm$ 0.17 &  5 & Nd  &    0.02 $\pm$ 0.13  & 5 \\
Cr  &  $-$0.12 $\pm$ 0.10 & 17 & Sm  &    0.16 $\pm$ 0.14  & 4 \\
\hline                                                                                                                                                          
\end{tabular}
\end{table}

\section{Time-series analysis and period change}
\label{sect:phot}
\begin{figure}
\centering
\includegraphics[width=8.5cm]{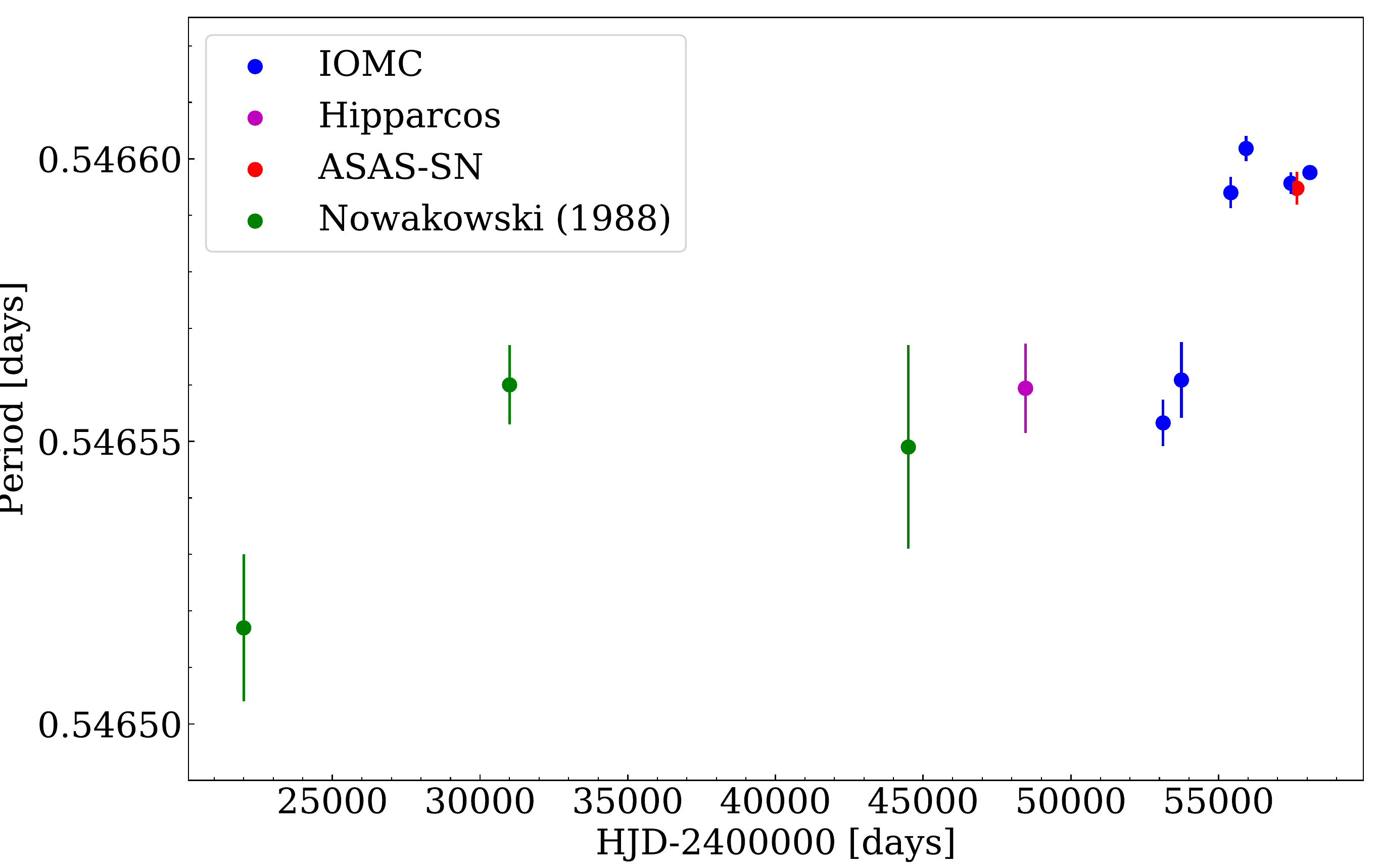}
\caption{Period change in the dominant pulsation mode (P1) of V363\,Cas. Labels identify the sources of different  literature datasets available for this star.}
\label{fig:pchange}
\end{figure}

The light-curve of V363\,Cas was studied in detail by \citet{Hajdu2009} who identified the true nature of DCEP of this source on the basis of the time-series data collected with the Integral Optical Monitoring Camera \citep[IOMC]{Alfonso2012} available at the time (1120 epochs over 6 years of data). However, since the IOMC has continued to accumulate data, we  retrieve all data available so far forV363\,Cas and analyse them again. At present the IOMC dataset comprises 3661 epochs with HJD spanning the range 2452654-2458504 days, (that is, from January 2003 to January 2019). This excellently long  time-series was analysed  with the Period04 period search software  \citep{Lenz2005}. We first considered the whole time-series, obtaining approximately the same pulsation periods as found by \citet{Hajdu2009}, but also detecting in the periodogram highly significant residual peaks around these values, a clear indication that the periods are not stable. We then subdivided the IOMC timeseries in six chunks covering approximately the same time span/number of epochs, to be able to derive the periods with sufficient precision and find possible changes. The result of this exercise is presented in Fig.~\ref{fig:pchange} (blue filled circles) for the P1 pulsation mode, as the P2 mode has a too small amplitude to allow precise results. A clear  change in period is seen along the 16 years spanned by the IOMC data, with a quick increase at HJD$\sim$2453000 days. To confirm these results, we searched the literature for additional time-series data, finding usable datasets from the Hipparcos mission \citep{Perryman1997} and the All-Sky Automated Survey for Supernovae  
%\citep[All-Sky Automated Survey for Supernovae]{Jayasinghe2018} 
\citep[ASAS-SN]{Jayasinghe2018}
surveys. Results from these datasets are plotted in red and cyan in Fig.~\ref{fig:pchange}. Additional insight into  the V363\,Cas  period changes were finally found in \citet{Nowakowski1988}, who analysed historical series of maxima for this star, providing periods valid in different epochs (green filled circles in Fig.~\ref{fig:pchange}). %with green filled circles. 
We can thus conclude that V363\,Cas is actually changing period, although at a rather slow rate, since overall its dominant period increased from P1=0.546517$\pm$0.000013 days to P1=0.546597$\pm$0.000001 days, i.e. 0.00008 days in 100 years, or $\dot{P}/P \sim 1.5\times10^{-6}$ year$^{-1}$. However, 
%as noted above, 
after HJD$\sim$2453000 days, the star had a sharp period increase corresponding to $\dot{P}/P \sim 1.0\times10^{-5}$ year$^{-1}$. This general trend is confirmed by the O--C data-set of V363\,Cas in the GEOS RR Lyr database \citep{Leborgne2007}.
%We can conclude that as expected, the increasing period of V363\,Cas shows that this star is at its first crossing as suggested by its position on the HR diagram.

\begin{figure}
\centering
\includegraphics[width=8.0cm]{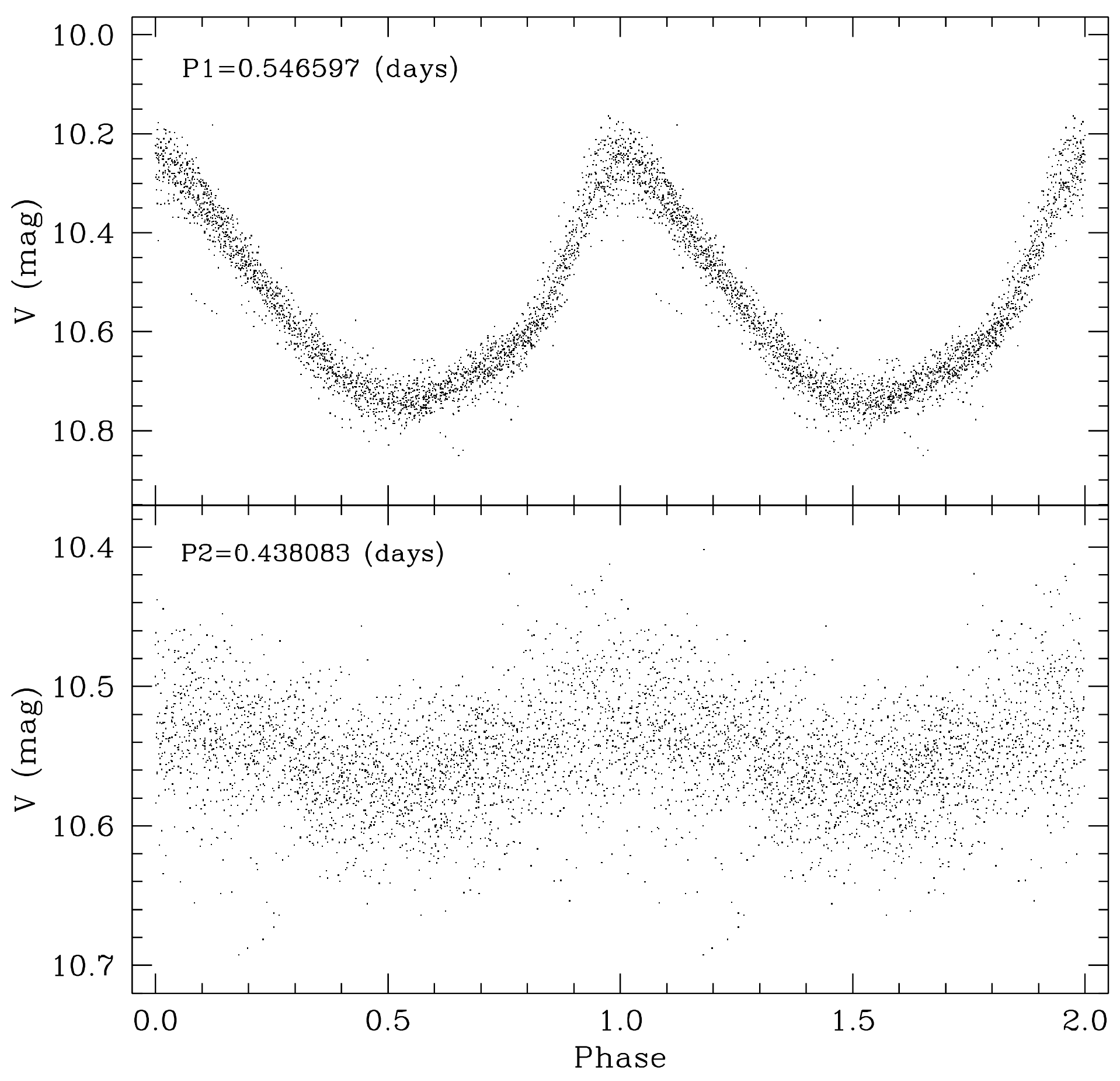}
\caption{Light curves of  V363\,Cas using  only  the IOMC dataset  with HJD in the range of 2455202-2458504 days. Top and bottom panels show the 1O (P1) and 2O (P2) light curves, respectively.}
\label{fig:lc}
\end{figure}

To provide updated periods and moments of maximum light for V363\,Cas, we performed a period search using only the four most recent chunks of the IOMC time-series dataset, as the periods inferred from them are found to be consistent to each other. Results from this procedure are summarised  in Table~\ref{tab:periods} whereas the folded light curves are shown in Fig.~\ref{fig:lc}.

\section{Discussion and conclusions}
\label{sect:discus}
\begin{figure}
\centering
\includegraphics[width=8.0cm]{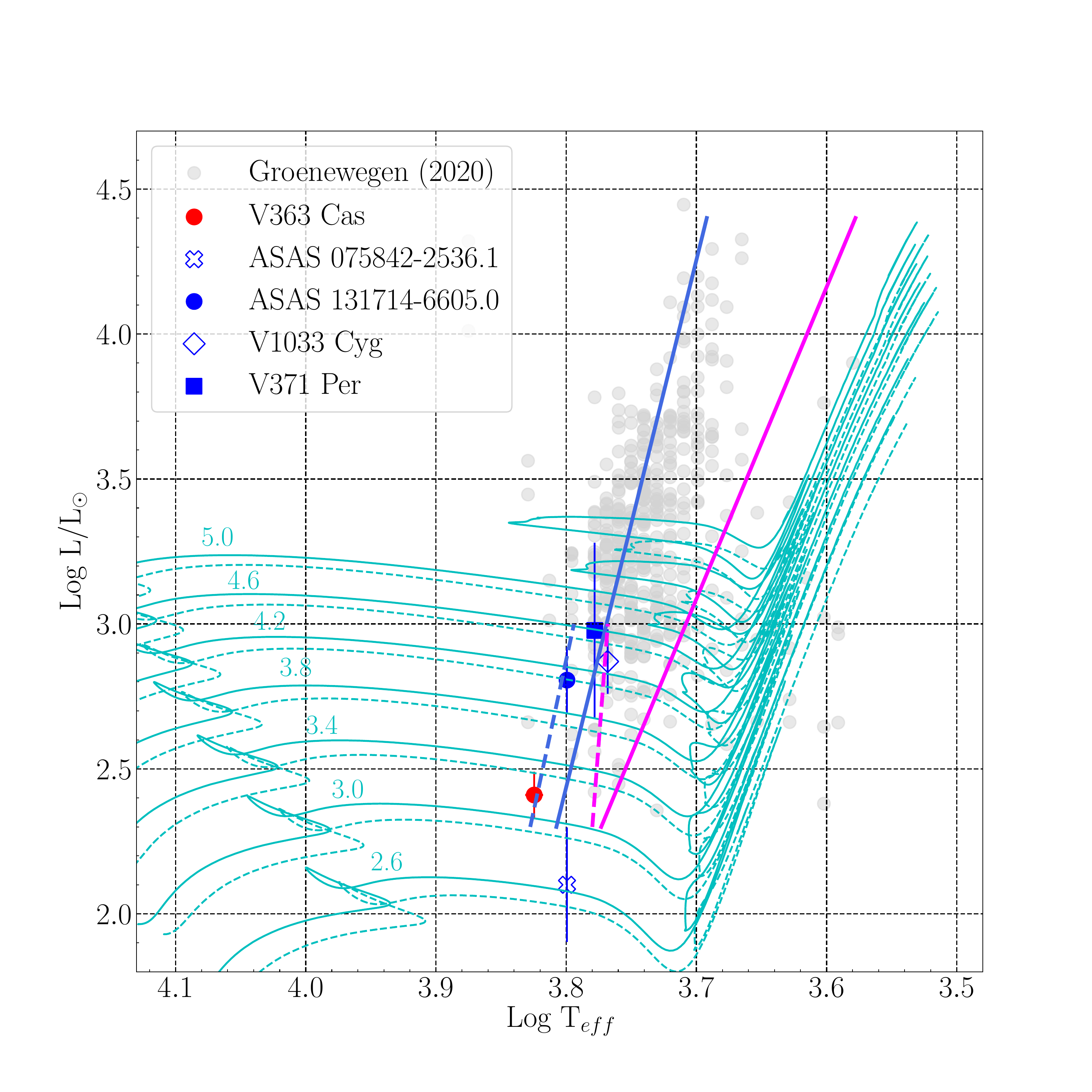}
\caption{HR diagram of the known MW Li-rich DCEPs. V363\,Cas is the red filled circle.  Different symbols (see labels) are used for the other 4  literature stars. The Li-normal DCEPs by \citet{Groenewegen2020} are shown in light grey as a reference. Instability strips for DCEP\_F (solid lines) and DCEP\_1O (dashed lines) by \citet{Desomma2020} as well as evolutionary tracks by \citet{Bressan2012} for Z=0.01 Y=0.267 (solid cyan lines) and Z=0.014 Y=0.273 (dashed cyan lines) in the mass range 2.6-5.0$M_{\odot}$ are also over-plotted to the data.}
\label{fig:hr}
\end{figure}

%Our spectroscopic analysis, in conjunction with the distances provided by {\it Gaia} mission and the integration of the SED to derive its 
The bolometric luminosity and effective temperature derived in our analysis, allows us to place V363\,Cas on the HRD. This is shown in Fig.~\ref{fig:hr}, where  we have also plotted,  for a comparison, the other four MW lithium-rich DCEPs.
%known to be lithium-rich from the literature. 
We adopted the $T_{\rm eff}$ values from \citet{Kovtyukh2019} for ASAS J075842-2536.1, ASAS J131714-6605.0 and V371 Per, while for V1033 Cyg we adopted the measure by \citet{Udovichenko2019}. The luminosities were estimated by integrating the SEDs, which, as for V363\,Cas, were calculated using the VOSA software.
%The adopted reddening of V1033\,Cyg $E(B-V)$=1.0$\pm0.1$ mag was taken from the compilation by \citet{Groenewegen2018}. For V371\,Per we used $E(B-V)$=0.18$\pm$0.02 obtained by converting the original value $E(V-R)$=0.136 by \citep{Schmidt2011} into $E(B-V)$ \citet{Cardelli1989} extinction law.
%No literature reddening value is available for ASAS J075842-2536.1 and V356 Mus. Therefore we decided to adopt the Period-Color relation for DCEP\_F by \citet{Tammann2003} (their Eq. 5) to estimate the intrinsic $(V-I)_0$ color. Therefore we have first to fundamentalize the 1O period of the two objects using the relation $P_F = P_{1O}/(0.716-0.027 \log P_{1O})$, being $P_F$  and $P_{1O}$ the periods of DCEP\_F and DCEP\_1O, respectively \citep[][]{Feast1997}. Secondly, we must obtain the apparent $(V-I)$ color, that was available in the literature. Hence we used the {\it Gaia} $(G_{BP}-G_{RP})$ color and transformed it into $(V-I)$ according to the conversion equations listed in the {\it Gaia} Data Release 2 Documentation (Sect. 5.3.7). Finally we obtained $E(B−V)=0.126\pm0.07$ and $E(B−V)=0.648\pm0.07$ for ASAS J075842-2536.1 and V356 Mus, respectively.  
Figure~\ref{fig:hr} also displays in light grey the location of the sample of Galactic DCEPs recently analysed by \citet{Groenewegen2020} together with the instability strip for DCEP\_F and DCEP\_1O of  \citet{Desomma2020}. Also plotted in  Fig.~\ref{fig:hr} are the   evolutionary tracks by \citet{Bressan2012}  in the mass range M=2.6-5.0 M$_{\odot}$ for the chemical composition  Z=0.01, Y=0.267 (which is adequate for 
%the estimated abundances of 
V363\,Cas and V371\,Per) and Z=0.014 Y=0.273 (close to the present solar metal content Z$_{\odot}$=0.0152 and suitable for the remaining stars). 
%An inspection of the figure reveals that
V363\,Cas is compatible with a mass of $\approx$\,3.2\,M$_\odot$ and is located at the blue side of the IS, as expected for a multi-mode DCEP pulsating in the 1O and 2O modes. In addition, the absence of extended  blue-loops in the tracks encompassing the position of V363\,Cas, supports our suggestion that V363\,Cas should be at its first crossing of the IS. %instability strip. 

\citet{takeda13} carried out a spectroscopic study of 12 evolved DCEPs to the aim of investigating 
%the nature of
evolution-induced mixing in the envelope of these stars. They derived photospheric abundances of C ($\approx -$0.30), N ($\approx$\,0.4-0.5), O (solar) and Na ($\approx$\,0.2). Moreover, Li has been found to be very low in DCEPs \citep[A(Li)<0.12][]{luck11}. V363\,Cas shows almost solar values for both  carbon and oxygen (no lines of N were detected in our spectral range) and a slight underabundance of sodium (but still consistent with the solar value within the experimental errors). Regarding lithium, by spectral synthesis of the resonance line at 6707.766 {\AA}, we derived A(Li)\,=\,2.86\,$\pm$\,0.10. This is consistent with the cosmic abundance and, at least, $\approx$\,1.8 dex over the average value found for DCEPs.

In summary, V363\,Cas seems to possess all characteristic features  of a first-crossing DCEP: it is  Li-rich, has solar-like C(N)O abundances (reflecting the photospheric abundance during the MS phase) and shows an increasing period. 
%We can compare V363\,Cas with the other known MW Li-rich DCEPs. 
The multi-mode nature of V363\,Cas is also a common feature among the few known Li-rich MW DCEPs.
%to the majority of Li-rich objects. 
Indeed, adding V363\,Cas, 4 out of 5 known MW Li-rich DCEPs are found to pulsate in two modes. Additionally, 3 are 2O/1O pulsators whereas only one is a 1O/F (V371\,Per) pulsator. In this respect, our results support \citet{Kovtyukh2019}  hypothesis that inefficient  mixing in the atmospheres of these DCEPs may have facilitated the preservation of Li. 
Regarding the position of the MW Li-rich DCEPs on the HRD, ASAS\,075852-2536.1 appears to be a low-mass DCEP ($\approx$\,2.6\,M$_\odot$), V1033\,Cyg is more massive ($\approx$\,4.4\,M$_\odot$) and is the only one which was shown to have an increasing period as we now  have found also of  V363\,Cas. V371\,Per ($\approx$\,4.6\,M$_\odot$) and ASAS\,311714-6605.0  ($\approx$\,4.2\,M$_\odot$) are far from the blue loops according to their tracks. Hence, all these DCEPs are very likely  at their first-crossing of the IS.

\begin{table}
\centering
      \caption{Periods, peak-to-peak amplitudes in the $V$ band and moment of maximum light ($-$2400000 days) of V363\,Cas.}
         \label{tab:periods}
\begin{tabular}{cccc}
\hline \hline
            \noalign{\smallskip}
  \multicolumn{1}{c}{Mode} &
  \multicolumn{1}{c}{Period} &
  \multicolumn{1}{c}{Amp(V)} &
  \multicolumn{1}{c}{HJD$_{Max}$} \\
            \noalign{\smallskip}
   \multicolumn{1}{c}{} &
  \multicolumn{1}{c}{days} &
  \multicolumn{1}{c}{mag} &
  \multicolumn{1}{c}{days} \\
           \noalign{\smallskip}
\hline
            \noalign{\smallskip}
P1 & 0.546597(1) & 0.444 & 55201.435 \\
P2 &  0.438083(2) & 0.046 & 55201.57 \\
P2/P1 & 0.80133(1) &  &  \\
            \noalign{\smallskip}
\hline\end{tabular}
\end{table}

To conclude, the emerging scenario is that a Li-rich DCEP should necessarily be at its first crossing (to avoid mixing due to 1DUP) and should have had a low rotation velocity when on the MS \citep[to avoid dilution by rotation mixing, see][]{Brott2011}. 
New high-resolution observations of MW DCEPs  from our and other groups are ongoing. Therefore,   more Li-rich DCEPs are likely to be discovered in the near future,  increasing the statistics and helping to unveil the true nature of these stars.

%The preferred mechanism for decreasing the angular momentum of the stars is the mass loss caused by stellar winds. According to the calculations carried out by \citet{krticka14}, we showed as this mechanism is not efficient in our case, since stars with masses lower than 5 M$_\odot$ do not have (or slight have) mass loss due to stellar wind, while massive stars reach the first crossing phase in a time of the order of 10$^7$ years, and than mass loss is negligible. As a consequence, these stars can not decrease angular momentum and their rotational velocity should remain almost constant during this evolutionary phases.

\begin{acknowledgements}

This work has made use of data from the European Space Agency (ESA) mission
{\it Gaia} (\url{https://www.cosmos.esa.int/gaia}), processed by the {\it Gaia} Data Processing and Analysis Consortium (DPAC,
\url{https://www.cosmos.esa.int/web/gaia/dpac/consortium}). Funding for the DPAC has been provided by national institutions, in particular the institutions participating in the {\it Gaia} Multilateral Agreement.
The Italian participation in DPAC has been supported by Istituto Nazionale di Astrofisica (INAF) and the Agenzia Spaziale Italiana (ASI) through grants I/037/08/0, I/058/10/0, 2014-025-R.0, and 2014-025-R.1.2015 to INAF (PI M.G. Lattanzi).
We acknowledge partial support from the project "MITiC: MIning The Cosmos Big Data and Innovative Italian Technology for Frontier Astrophysics and Cosmology”  (PI B. Garilli).
This publication makes use of VOSA, developed under the Spanish Virtual Observatory project supported by the Spanish MINECO through grant AyA2017-84089.
VOSA has been partially updated by using funding from the European Union's Horizon 2020 Research and Innovation Programme, under Grant Agreement nº 776403 (EXOPLANETS-A).
This research has made use of the SIMBAD database,
operated at CDS, Strasbourg, France.
\end{acknowledgements}

%\begin{appendix} %First appendix

%\end{appendix}


\begin{thebibliography}{}

\bibitem[Alfonso-Garz{\'o}n et al.(2012)]{Alfonso2012} Alfonso-Garz{\'o}n, J., Domingo, A., Mas-Hesse, J.~M., et al.\ 2012, \aap, 548, A79

\bibitem[Bayo et al.(2008)]{bayo08} Bayo, A., Rodrigo, C., Barrado y Navascués, D., Solano, E., Gutiérrez, R., Morales-Calderón, M., Allard, F. 2008, A\&A 492,277B.

\bibitem[Bressan et al.(2012)]{Bressan2012} Bressan, A., Marigo, P., Girardi, L., et al.\ 2012, \mnras, 427, 127

\bibitem[Brott et al.(2011)]{Brott2011} Brott, I., de Mink, S.~E., Cantiello, M., et al.\ 2011, \aap, 530, A115

\bibitem[Catanzaro et al. (2019)]{catanzaro19} Catanzaro, G., Bus\'a, I., Gangi M., Giarrusso, M., Leone, F., Munari, M., 2019, \mnras, 484, 2530

\bibitem[Cyburt et al.(2008)]{Cyburt2008} Cyburt R. H., Fields B. D., Olive K. A., 2008, J. Cosmol. Astropart. Phys., 11, 12

\bibitem[Dambis et al.(2013)]{Dambis2013} Dambis, A.~K., Berdnikov, L.~N., Kniazev, A.~Y., et al.\ 2013, \mnras, 435, 3206

\bibitem[De Somma et al.(2020)]{Desomma2020} De Somma, G., Marconi, M., Molinaro, R., et al.\ 2020, \apjs, 247, 30

\bibitem[Fernley \& Barnes (1997)]{fernley97}  Fernley, J., \& Barnes, T.~G. 1997, \aaps, 125, 313

\bibitem[Fitzpatrick (1999)]{fitz1999} Fitzpatrick, E., 1999, PASP, 111, 63

\bibitem[Grevesse et al. (2010)]{grevesse10} Grevesse N., Asplund M., Sauval A. J., Scott P., 2010, Ap\&SS, 328, 179

\bibitem[Groenewegen(2018)]{Groenewegen2018} Groenewegen, M.~A.~T.\ 2018, \aap, 619, A8

\bibitem[Groenewegen(2020)]{Groenewegen2020} Groenewegen, M.~A.~T.\ 2020, \aap, 635, A33

\bibitem[Hajdu et al.(2009)]{Hajdu2009} Hajdu, G., Jurcsik, J., \& Sodor, A.\ 2009, Information Bulletin on Variable Stars, 5882, 1

\bibitem[Huang et al.(2010)]{Huang2010} Huang, W., Gies, D.~R., \& McSwain, M.~V.\ 2010, \apj, 722, 605

\bibitem[Iben(1967)]{Iben1967} Iben, I.\ 1967, \araa, 5, 571


%{\bf \bibitem[Indebetouw et al. (2005)]{inde2005} Indebetouw, R., Mathis, J. S., Babler, B. L.,             et al., 2005, ApJ 619, 931  Serve davvero questa referenza??}

\bibitem[Jayasinghe et al.(2018)]{Jayasinghe2018} Jayasinghe, T., Kochanek, C.~S., Stanek, K.~Z., et al.\ 2018, \mnras, 477, 3145

\bibitem[Kervella et al.(2019a)]{Kervella2019a} Kervella, P., Gallenne, A., Evans, N.~R., et al.\ 2019a, \aap, 623, A117

\bibitem[Kervella et al.(2019b)]{Kervella2019b} Kervella, P., Arenou, F., Mignard, F., et al.\ 2019b, \aap, 623, A72

\bibitem[Kovtyukh \& Gorlova (2000)]{Kovtyukh2000} Kovtyukh, V.~V., Gorlova, N.~I. \ 2000, \aap, 358, 587

\bibitem[Kovtyukh et al.(2019)]{Kovtyukh2019} Kovtyukh, V., Lemasle, B., Kniazev, A., et al.\ 2019, \mnras, 488, 3211

\bibitem[Kovtyukh et al.(2016)]{Kovtyukh2016} Kovtyukh, V., Lemasle, B., Chekhonadskikh, F., et al.\ 2016, \mnras, 460, 2077

%\bibitem[Krti\v{c}ka (2014)]{krticka14} Krti\v{c}ka, J., 2014,  \aap 564, A70

\bibitem[Kurucz (1993a)]{kur93}  Kurucz R.L., 1993, A new opacity-sampling model atmosphere  program for arbitrary abundances. In: Peculiar versus normal phenomena in 
 A-type and related stars, IAU Colloquium 138, M.M. Dworetsky, F. Castelli, 
 R. Faraggiana (eds.), A.S.P Conferences Series Vol. 44, p.87
 
\bibitem[Kurucz (1993b)]{kur93b} Kurucz R.L., 1993, Kurucz CD-ROM 13: ATLAS9, SAO, Cambridge, USA 

\bibitem[Kurucz \& Avrett (1981)]{kur81} Kurucz R.L., Avrett E.H., 1981, SAO Special Rep., 391 

\bibitem[Le Borgne et al.(2007)]{Leborgne2007} Le Borgne, J.~F., Paschke, A., Vandenbroere, J., et al.\ 2007, \aap, 476, 307

\bibitem[Lenz \& Breger(2005)]{Lenz2005} Lenz, P., \& Breger, M.\ 2005, Communications in Asteroseismology, 146, 53

\bibitem[Luck \& Lambert (2011)]{luck11} Luck, R. E., Lambert, D. L., 2011, \aj, 142, 136 

\bibitem[Luck \& Lambert (1992)]{luck92} Luck, R. E., Lambert, D. L., 1992, \apjs, 79, 303 

\bibitem[Maeder \& Meynet(2000)]{Maeder2000} Maeder, A., \& Meynet, G.\ 2000, \araa, 38, 143

\bibitem[Nowakowski(1988)]{Nowakowski1988} Nowakowski J. 1988, J. Am. Assoc. Variable star obs., 17, 7 

\bibitem[Perryman et al.(1997)]{Perryman1997} Perryman, M.~A.~C., Lindegren, L., Kovalevsky, J., et al.\ 1997, \aap, 500, 501


\bibitem[Prudil et al.(2020)]{Prudil2020} Prudil, Z., D{\'e}k{\'a}ny, I., Grebel, E.~K., et al.\ 2020, \mnras, 492, 3408

\bibitem[Romaniello et al.(2008)]{Romaniello2008} Romaniello, M., Primas, F., Mottini, M., et al.\ 2008, \aap, 488, 731

\bibitem[Takeda et al. (2013)]{takeda13} Takeda, Y., Kang, D.-I., Han, I., Lee3, B.-C., \&  Kim, K.-M., 2013, \mnras, 432, 769

\bibitem[Turner et al.(2006)]{Turner2006} Turner, D.~G., Abdel-Sabour Abdel-Latif, M., \& Berdnikov, L.~N.\ 2006, \pasp, 118, 410

\bibitem[Udalski et al.(2018)]{Udalski2018} Udalski, A., Soszy{\'n}ski, I., Pietrukowicz, P., et al.\ 2018, \actaa, 68, 315

\bibitem[Udovichenko et al.(2019)]{Udovichenko2019} Udovichenko, S.~N., Kovtyukh, V.~V., \& Keir, L.~E.\ 2019, Odessa Astronomical Publications, 32, 83


\end{thebibliography}
\end{document}